# Magnetization, Structural and Nuclear Quadrupole Resonance Study of $RuSr_2EuCeCu_2O_{10+d}$.

G. V. M. Williams,

*2. Physikalisches Institut, Universität Stuttgart, D-70550 Stuttgart, Germany*

*and Industrial Research Limited, P.O. Box 31310, Lower Hutt, New Zealand.*

A. C. McLaughlin and J. P. Attfield

*IRC in Superconductivity, University of Cambridge, Madingley Road, Cambridge CB3 0HE, United Kingdom.*

S. Krämer,

*2. Physikalisches Institut, Universität Stuttgart, D-70550 Stuttgart, Germany.*

Ho Keun Lee.

*Department of Physics, Kangwon National University, Chunchon 200-701, Republic of Korea.*

**ABSTRACT**

We present the results from a magnetization, structural, and nuclear quadrupole resonance study of the ruthenate-cuprate, $RuSr_2R_{2-x}Ce_xCu_2O_{10+\delta}$ with x=1 and R=Eu. This compound is a superconductor for Ce doping in the range $0.4 \leq x \leq 0.8$ and displays ferromagnetic order for $0.4 \leq x \leq 1.0$. We show that the tilting and rotation of the $RuO_6$ octahedra are essentially the same for x=1, x=0.6 and the superconducting antiferromagnet, $RuSr_2GdCu_2O_8$. However, the moment per Ru in $RuSr_2EuCeCu_2O_{10+\delta}$ is comparable to that observed in $RuSr_2EuCu_2O_8$. These results indicate that it is unlikely that the different magnetic order found in $RuSr_2R_{2-x}Ce_xCu_2O_{10+\delta}$ and $RuSr_2RCu_2O_8$ (R=Y,Gd,Eu) is due to changes in structurally induced distortions of the $RuO_6$ octahedra as found in the ruthenate compounds (e.g. $Sr_{1-x}Ca_xRuO_3$). We show that the Cu nuclear quadrupole resonance (NQR) data are similar to those observed in the single $CuO_2$ layer superconductor, $La_{1-x}Sr_xCuO_4$, where the Cu spin-lattice relaxation rate is dominated by hyperfine coupling within the $CuO_2$ layers and any additional hyperfine coupling from spin fluctuations in the $RuO_2$ layers is small.

**Introduction**

The ruthenate-cuprates ($RuSr_2R_{2-x}Ce_xCu_2O_{10+\delta}$ and $RuSr_2RCu_2O_8$, R=Y,Eu,Gd) are proving to be particularly interesting because they display magnetic order that is not seen in other ruthenate compounds (e.g. $Sr_{1-x}Ca_xRuO_3$ and $Sr_2RRuO_6$). These compounds were discovered by Bauernfeind, Widder and Braun [1,2] and they have been found to display weak ferromagnetic ($RuSr_2R_{2-x}Ce_xCu_2O_{10+\delta}$ [3]) or predominately antiferromagnetic ($RuSr_2RCu_2O_8$ [4,5,6]) order as well as the coexistence of superconductivity [7,8,9]. The ruthenate-cuprates contain $CuO_2$ layers as well as $RuO_2$ layers that magnetically order. A detailed measurement and characterization of the macroscopic and microscopic properties of the $RuO_2$ layers, the $CuO_2$ layers and their mutual interactions is a key ingredient for a consistent theoretical description of these materials.

Recent x-ray absorption near edge spectroscopy (XANES) studies on $RuSr_2R_{2-x}Ce_xCu_2O_{10+\delta}$ and $RuSr_2RCu_2O_8$ have indicated that the average Ru valence is 4.6 [10] for $RuSr_2RCu_2O_8$ and 5.0 [11] or 4.95 [12] for $RuSr_2R_{2-x}Ce_xCu_2O_{10+\delta}$. The appearance of a mixed Ru valence in the case of $RuSr_2RCu_2O_8$ has also been observed by nuclear magnetic resonance (NMR) measurements [13,14]. It is surprising that a mixed Ru valence is observed in $RuSr_2RCu_2O_8$ because a mixed Ru valence is not expected in, and has not been reported in, other ruthenate compounds. This may suggest that these materials display an intrinsic inhomogeneous distribution of charge and spin density. Furthermore, the Ru valence in $RuSr_2R_{2-x}Ce_xCu_2O_{10+\delta}$ is nearly the same as that from the antiferromagnetic insulator, $Sr_2RCuO_6$. Thus, there seems to be no direct correlation between the valence state of Ru and the observed magnetic ordering state, suggesting that more complicated models involving other physical parameters might be required.

There is also the question of the effect of Ce doping in $RuSr_2R_{2-x}Ce_xCu_2O_{10+\delta}$. It has been found that there is no significant change in $T_c$ for $0.4 \leq x \leq 0.8$ indicating that there is no change in the hole concentration on the $CuO_2$ planes [15]. Furthermore, there is no significant charge transfer to the $RuO_2$ planes for $0.5 \leq x \leq 1.0$ [12]. It is believed, based on x-ray absorption spectroscopy measurements on a related compound ($Pr_{1.5}Ce_{0.5}NbSr_2Cu_2O_{10+\delta}$ [16]), that the Ce valence in $RuSr_2R_{2-x}Ce_xCu_2O_{10+\delta}$ is 4+ [11]. Thus, if all the carriers introduced by Ce doping appeared in the $CuO_2$ planes then it would be expected that an increase in the Ce concentration from 0.4 to 0.8 should correspond to a decrease in the hole concentration on the $CuO_2$ planes of 0.2. This change in hole concentration is large and it is comparable to the hole concentration range where superconductivity is observed in the high temperature superconducting cuprates (HTSC) ($0.05 \leq p \leq 0.27$). It has been suggested that the carriers introduced by Ce are partially compensated by an increase in the oxygen content [15].

It is not clear if the different magnetic order in the ruthenate-cuprates can be explained by the current theories that have been applied to $Sr_{1-x}Ca_xRuO_3$, which also has $RuO_6$ octahedra. In particular, the different magnetic order observed in $Sr_{1-x}Ca_xRuO_3$ is believed to be due to changes in the bandwidth induced by the changes in the Ru-O-Ru bond lengths as well as tilting and rotations of the $RuO_6$ octahedra [17]. It is known that a rotation and tilting of the $RuO_6$ octahedra also occurs in $RuSr_2GdCu_2O_8$ where the rotations form coherent domains extending up to 20 nm and results in super-cell features [18]. One neutron powder diffraction study has shown that a rotation and tilting of the $RuO_6$ octrahedra also occurs in $RuSr_2Gd_{1.4}Ce_{0.6}Cu_2O_{10+\delta}$ [19]. However, it is not known if they occur for other Ce doping concentrations. This is especially pertinent because it is known that the magnetization at 1 T and the peak in the temperature-dependent magnetization data increase with increasing Ce concentration [15].

In this paper, we report the results from a magnetization, synchrotron x-ray diffraction and nuclear quadrupole resonance (NQR) study of $RuSr_2EuCeCu_2O_{10+\delta}$ with the aim to address the problems above. The techniques applied in this study provide insight into the macroscopic properties as probed by magnetization measurements as well as microscopic information concerning the atomic positions and bond angles or the local static and dynamic properties as seen by the $^{63,65}Cu$ nuclei. We show that this compound is a ferromagnet with a saturation moment at 6 T that is comparable to that observed in the antiferromagnetic superconductor, $RuSr_2EuCu_2O_8$. Furthermore, the tilting and rotation angles are comparable to those observed in superconducting $RuSr_2GdCu_2O_8$ and $RuSr_2Gd_{1.4}Ce_{0.6}Cu_2O_{10+\delta}$. We show that the NQR spin-lattice relaxation rate is dominated by coupling within the $CuO_2$ planes and the Cu spin-lattice relaxation rate is comparable to that of the single $CuO_2$ layer HTSC, $La_{2-x}Sr_xCuO_4$.

**Experimental Details**

The $RuSr_2EuCeCu_2O_{10+\delta}$ ceramic sample was made from a stoichiometric mix of $RuO_2$, $SrCO_3$, $Eu_2O_3$, $CeO_2$ and CuO. The material was decomposed at 960 °C for 1 hour and then pressed into pellets. The pellets were placed in flowing $N_2$ for 10 hours, after which the material was ground, pressed into pellets and placed in flowing $O_2$ for 10 hours at 1070 °C, 24 hours at 1075 °C, 24 hours at 1080 °C, 24 hours at 1080 °C and 6 days at 1080 °C, with intermediate grinding. The sample was removed for the hot zone and then furnace-cooled to 700 °C after which the sample was reinserted and slow cooled to room temperature. There was no evidence of impurity phases in the synchrotron x-ray spectra. A $RuSr_2EuCu_2O_8$ sample was prepared using the synthesis method described elsewhere [2,8]. X-ray diffraction measurements indicated that the sample was single phase. It is not possible to c-axis align $RuSr_2EuCeCu_2O_{10+\delta}$ or $RuSr_2EuCu_2O_8$ and hence the experiments were performed on unoriented ceramic samples or powders.

Synchrotron X-ray diffraction patterns were recorded at Station 9.1, Daresbury SRS (synchrotron radiation source), UK, which is a high-energy, high resolution monochromatic diffraction station for structure solution, time resolved and high pressure studies. The sample was placed in a 0.5 mm diameter borosilicate glass capillary mounted on the axis of the diffractometer about which it was spun at a frequency of approximately 1Hz to improve the powder averaging of the crystallites. Diffraction patterns were collected at room temperature with a step size of 0.01° and, in order to minimize absorption, a wavelength of 0.805521 Å was used. The high-angle parts of the pattern were scanned several times to improve the statistical quality of the data in these regions. Finally the counts from the detectors were normalized, summed and rebinned; the patterns were then refined by the Rietveld method using the GSAS program [20]. The full width at half maximum for this instrument in Debye – Scherrer geometry is ~ 0.03 – 0.05º and the peaks are easily described by a Pseudo Voigt function.

The ac and dc magnetization measurements were made using a SQUID magnetometer. The remnant magnetic field from the superconducting magnet was ±0.6 mT. The ac magnetization data was taken with an ac magnetic field of 5 µT and a frequency of 100 Hz. Room temperature thermopower measurements were made using the standard differential temperature technique. Electrical resistance measurements were made between 4 K and 300 K using the four terminal technique.

Cu NQR measurements were performed on a powder sample using a home built pulsed NQR spectrometer, described elsewhere [21]. The probe was placed in a continuous flow cryostat, which was encased in a µ-metal shield in order to reduce the magnetic field at the sample from stray magnetic fields and the earth's magnetic field. An inversion recovery pulse sequence was used to measure the spin-lattice relaxation rate, $1/T_1$, and a Hahn echo sequence was used to obtain the NQR spectra. In both cases the π/2 pulse width was 1 µs and the time between the π/2 pulse and the π pulse was 9-10 µs. The NQR spectra were obtained at discrete frequency steps and then adding the Fourier transformed spectra from the second half of the echo after phase correction.

In homogeneous materials the Cu NQR spin-lattice magnetization recovery for the inversion recovery pulse sequence is a mono-exponential and can be expressed as,

$$M(t) = M_0 \left(1 - 2\exp\left(-\frac{3t}{T_1}\right)\right), \quad (1)$$

where $\tau$ is the time between the first inversion pulse and the first pulse of the detection echo sequence. The additional factor of 3 is added in order to enable a direct comparison between the NQR and nuclear magnetic resonance (NMR) spin-lattice relaxation rates.

**Results and Analyses**

The unit cell of $RuSr_2R_{2-x}Ce_xCu_2O_{10+\delta}$ is plotted in figure 1. It contains subunits with two $CuO_2$ layers that are separated by two RO layers. This defines a subunit that has similar dimensions and atomic positions as the single layer electron-doped HTSC, $R_{2-x}Ce_xCu_2O_4$. The unit cell also contains $RuO_2$ layers that are separated from the $CuO_2$ layers by an insulating SrO layer. The unit cell and atomic positions are similar to those found in $NbSr_2Eu_{2-x}Ce_xCu_2O_{10}$ [22]. However, we show later that $RuSr_2RCeCu_2O_{10+\delta}$ also contains a rotation and tilt of the $RuO_6$ octahedra similar to those found in $RuSr_2Gd_{1.4}Ce_{0.6}Cu_2O_{10+\delta}$ [19] and $RuSr_2GdCu_2O_8$ [18]. The tilt and rotation of the $RuO_6$ octahedra are indicated in figure 1 by the multiple O(3) and O(1) positions.

The room temperature thermopower for the x=1 and R=Eu sample is 69 µV/K. A comparison with the room temperature thermopower values of the double $CuO_2$ layer HTSC [23] would indicate that the hole concentration of this sample is ~0.05 and it is on the insulator-superconductor boundary. However, the results from a recent study on single $CuO_2$ layer $La_{2-x}Sr_xCuO_4$ would appear to suggest that the hole concentration is near 0.075 [24]. There is no evidence of superconductivity from the electrical resistance data (not shown) and the resistance increases with decreasing temperature.

There is also no evidence of superconductivity or a superconducting fraction in the ac susceptibility data plotted in figure 2. Rather, the main feature seen in the real part of the ac susceptibility, $c'$ (solid curve, left axis), is a peak near 106 K which is associated with the magnetic ordering transition. The peak in the imaginary part of the ac susceptibility, $c''$, occurs at a slightly lower temperature of 100 K as can be seen in figure 2b. Similar peaks are seen in $RuSr_2RCu_2O_8$ but at a higher temperature of 132 K [8,25,26], as can be seen in figure 2a (dashed curve, right axis) for R=Eu. As mentioned above, the low-field magnetic order in $RuSr_2RCu_2O_8$ is predominately antiferromagnetic and the peak in $c'$ occurs at essentially the same temperature as the magnetic ordering temperature [25], defined as temperature where $dM/dT$ is a maximum negative value.

It is apparent in figure 3 that the low field magnetic ordering temperature in $RuSr_2EuCeCu_2O_{10+\delta}$ is 115 K, where we use the definition mentioned above. Here we plot the field-cooled and zero-field-cooled dc magnetization data from $RuSr_2EuCeCu_2O_{10+\delta}$ (solid curves, left axis) and $RuSr_2EuCu_2O_8$ (dashed curves, right axis) for an applied magnetic field of -2.5 mT. We find that the magnetic ordering temperature increases with increasing applied magnetic field as can be seen in the lower insert to figure 3, where the magnetic ordering temperature for an applied magnetic field of 6 T is 146 K. It can be seen that the peak in the zero-field-cooled dc magnetization, for an applied magnetic field of –2.5 mT, occurs at a lower temperature of 94 K. This can be contrasted with $RuSr_2RCu_2O_8$ where the peak in the dc magnetization for low applied fields occurs at essentially the same temperature as the temperature where $-dM/dT$ from the field-cooled data is a maximum [25]. This peak is due to reorientation and growth of ferromagnetic domains and shifts to lower temperatures with increasing magnetic fields. For low temperatures, the domain wall pinning energy is greater than $k_BT$. As the temperature is increased, the weakly pinned domain walls are free to move and hence the magnetization increases.

The higher magnetization observed in the field-cooled condition and for low applied magnetic fields and temperatures implies a magnetic irreversibility, which disappears for magnetic fields greater than ~ 0.2 T for $RuSr_2EuCeCu_2O_{10+\delta}$ but it is still evident for fields of up to 3 T in $RuSr_2RCu_2O_8$. For low applied magnetic fields, the irreversibility occurs at a temperature that is 10 K greater than the magnetic ordering temperature. This can be seen in the upper insert to figure 3 for $RuSr_2EuCeCu_2O_{10+\delta}$ (solid curve, left axis) and $RuSr_2EuCu_2O_8$ (dashed curve, right axis). We note that the irreversibility temperature is close to the temperature where the onset of ac loss occurs as can be seen in figure 2b.

It is apparent in figure 4 that the magnetic field dependence of the magnetization is different for $RuSr_2EuCeCu_2O_{10+\delta}$ and $RuSr_2EuCu_2O_8$. Here we plot the magnetization from $RuSr_2EuCeCu_2O_{10+\delta}$ (solid curve) and $RuSr_2EuCu_2O_8$ (dashed curve) at 5 K after subtracting the small van Vleck susceptibility from $Eu^{3+}$, which will be discussed later. In particular, $RuSr_2EuCu_2O_8$ has a small remnant magnetization of ~0.035 $\mu_B$/Ru, which is slightly smaller than that found in an earlier study [25], while $RuSr_2EuCeCu_2O_{10+\delta}$ has a remnant magnetization of 0.40 $\mu_B$/Ru. These values correspond to ~5 % and ~38 % of the respective magnetization values at 6 T (0.73 $\mu_B$/Ru for $RuSr_2EuCu_2O_8$ and 1.06 $\mu_B$/Ru for $RuSr_2EuCeCu_2O_{10+\delta}$). The small remnant magnetization in $RuSr_2RCu_2O_8$ has been attributed to a ferromagnetic component due to spin canting where the Ru moments are antiferromagnetically ordered in all three directions [5,6,25]. In the case of $RuSr_2GdCu_2O_8$, the remnant moment is significantly larger (0.14 $\mu_B$/Ru [8]) which has been attributed to coupling of the ferromagnetic component with the $Gd^{3+}$ moment [5]. The situation is more complicated because, although a recent theoretical study showed that the antiferromagnetic state is energetically favored over the antiferromagnetic state [27], a recent powder neutron diffraction study on $RuSr_2YCu_2O_8$ has found a significant ferromagnetic component of 0.28 $\mu_B$/Ru [28]. We note that it has been suggested that the high-field magnetic order in $RuSr_2EuCu_2O_8$ is ferromagnetic and hence there is a spin-flop transition [25].

The large remnant magnetization, in relation to the magnetization at 6 T, found in $RuSr_2EuCeCu_2O_{10+\delta}$ is consistent with ferromagnetic order although neutron diffraction measurements are required to precisely determine the nature of the magnetic order. Unlike $RuSr_2EuCu_2O_8$, the magnetization from $RuSr_2EuCeCu_2O_{10+\delta}$ rapidly increases for magnetic fields of up to ~0.2 T after which the increase in the magnetization is more gradual. Similar behavior is also observed in the itinerant ferromagnetic metals, $SrRuO_3$ ($T_m$=165 K [29]) and $Sr_4Ru_3O_{10}$ ($T_m$=148 K [30]) which have magnetization values at 6 T of ~1.25 $\mu_B$/Ru [29] and ~1.6 $\mu_B$/Ru [30] respectively. For both of these compounds, the magnetization saturates to the low-spin S=1 value expected for the $t_{2g}^4$ ground state where the saturation fields are ~40 T for $SrRuO_3$ [31] and ~16 T for $Sr_4Ru_3O_{10}$ [30]. It is surprising that the low field magnetization from $RuSr_2EuCeCu_2O_{10+\delta}$ is not higher than the observed 1.06 $\mu_B$/Ru at 6 T because recent studies have shown that the Ru valance is near 5+ [12]. Thus, the ground state should be $t_{2g}^3$ with S=3/2 and hence the saturation magnetization should be 3 $\mu_B$/Ru. We note that it has also been suggested that $RuSr_2Eu_{1.5}Ce_{0.5}Cu_2O_{10+\delta}$ is in a low spin state with S=1/2 in which case the saturation moment for $RuSr_2EuCeCu_2O_{10+\delta}$ should be 1 $\mu_B$/Ru [32].

Previous studies of the magnetic order in $SrRuO_3$ and $RuSr_2EuCu_2O_8$ have used the high temperature magnetization data to assist in the determination of the nature of the magnetic order as well as the Ru valence [29,33,34]. For this reason, we plot M/H against temperature for magnetic fields of 1 T, 3 T and 6 T in figure 5. It is apparent that spin fluctuation effects occur for temperatures below 200 K where the M/H curves separate. We note that this is similar to $RuSr_2EuCu_2O_8$ where spin fluctuation effects are evident below 200 K [25]. Therefore, we model the M/H data above 200 K and in the region where M/H is independent of applied magnetic field by $c(T) = c/(T-J) + c_{Eu}(T) + c_0$. The first term is the Curie-Weiss contribution from the Ru moment where $J$ is the Curie-Weiss temperature. The second term is the van Vleck susceptibility from $Eu^{3+}$ ($^7F_0$). Similar to the high temperature susceptibility study on $RuSr_2EuCu_2O_8$ [33], we use a spin-orbit coupling constant of 303 cm$^{-1}$ as measured in $Eu_2CuO_4$ [35]. However, as we show below, the deduced Ru moment is not critically dependent on the value of the spin-orbit coupling constant. The van Vleck susceptibility from $Eu^{3+}$ is plotted in figure 5 (dotted curve). Also shown is the susceptibility after subtraction of the $Eu^{3+}$ component (dashed curve). The third term accounts for the susceptibility from the $CuO_2$ planes and temperature independent susceptibility from the $RuO_2$ layers. The spin susceptibility from the $CuO_2$ planes in the HTSC is small ($<4\times10^{-5}$ [36]) and hence we approximate it as a constant for temperatures above 200 K. We show later that the NQR frequency from $RuSr_2EuCeCu_2O_{10+\delta}$ is comparable to that observed in the HTSC and hence there is no Cu moment contribution from the possible antiferromagnetic insulating state that is observed in very underdoped HTSC ($p<0.05$).

We show by the solid curve in figure 5 that the susceptibility, after subtracting the $Eu^{3+}$ contribution, can be fitted to a Curie-Weiss function and a constant term. We find that the effective high-temperature moment is 2.6 $\mu_B$/Ru, $q$ =133 K and $c_0$ =14×10$^{-5}$. Note that varying the spin-orbit coupling constant from 250 cm$^{-1}$ to 350 cm$^{-1}$ does not significantly change the fitted parameters. While the effective high-temperature moment is greater than that expected for a $t_{2g}^3$ ground state and S=1/2 (1.73 $\mu_B$/Ru), it is considerably less than that expected for a $t_{2g}^3$ ground state and S=3/2 (3.87 $\mu_B$/Ru). Furthermore, $q$ is 18 K greater than the magnetic ordering temperature. The fitted value of $c_0$ is significantly greater than $c(T)$ in $YBa_2Cu_3O_7$ which gives an upper limit to the Pauli spin susceptibility and the diamagnetic term from the $CuO_2$ planes of ~4×10$^{-5}$ [36]. The remaining $c_0$ is large but it is comparable to, or smaller than, the temperature independent susceptibility from $Sr_2YRuO_6$ and $SrRuO_3$ [34].

We show in figure 6 that the synchrotron x-ray spectra can be refined to a I4/mmm space group where the atomic positions are given in table 1 and the bond angles and bond lengths are given in table 2. It is clear from table 2 that the $RuO_6$ octahedra are tilted by 5.7° and rotated about the $c$-axis by 14.2°. It is remarkable that nearly the same tilt and rotation angles have also been observed in $RuSr_2Gd_{1.4}Ce_{0.6}Cu_2O_{10+\delta}$ [19] and $RuSr_2GdCu_2O_8$ [18] as can be seen in table 2. As mentioned above, the octahedral distortions in $RuSr_2GdCu_2O_8$ form coherent domains leading to super-cell features. There is no evidence for similar super-cell peaks in the current $RuSr_2EuCeCu_2O_{10+\delta}$ sample. Thus, although $RuSr_2Gd_{2-x}Ce_xCu_2O_{10+\delta}$ and $RuSr_2GdCu_2O_8$ have different magnetic order, we find that the tilting and rotation of the $RuO_6$ octahedra are similar. As mentioned earlier, changes in the magnetic order in other ruthenate compounds are normally accompanied by significant changes in the tilting and rotation angles of the $RuO_6$ octahedra.

It is interesting to note that the Ru-O-Cu bond length is essentially the same in $RuSr_2EuCeCu_2O_{10+\delta}$, $RuSr_2Gd_{1.4}Ce_{0.6}Cu_2O_{10+\delta}$ and $RuSr_2GdCu_2O_8$ but the apical oxygen (O(1)) is closer to Cu in $RuSr_2EuCeCu_2O_{10+\delta}$ as can be seen in table 2. Changes in the distance between the apical oxygen and the planar Cu have been observed in the HTSC, $YBa_2Cu_3O_{7-\delta}$, with increasing oxygen content, and have been interpreted in terms of charge transfer to the $CuO_2$ layers [37]. These changes are large (~0.16 Å) when compared with the corresponding change in the Cu to plane oxygen (O(2)) bond length (~0.01 Å) when the oxygen content is changed from 6 to 7. A much smaller increase in the Cu to apical oxygen bond length is observed in single layer $La_{2-x}Sr_xCuO_4$ (~0.02 Å) with increasing hole concentration [38]. However, the Cu-O(1) apical bond length in the ruthenate-cuprates (<2.21 Å) is much shorter than that observed in $YBa_2Cu_3O_{7-\delta}$ (>2.30 Å [37]), $La_{2-x}Sr_xCuO_4$ (2.42 Å [38]) as well as $TlSr_2(R_{1-x}Ca_x)Cu_2O_{7-\delta}$ (>2.30 Å [39]). In the case of $RuSr_2GdCu_2O_8$, bond valance sum analysis has been used to show that this may imply a higher hole concentration in the $CuO_2$ planes of p~0.4, which is much higher than that observed in the superconducting region of the HTSC (0.05<p<0.27) [18]. A similar bond valance sum analysis was performed on $RuSr_2Gd_{1.4}Ce_{0.6}Cu_2O_{10+\delta}$ where it was found that p also is near +0.4 [19]. However, the calculation assumes that Ce is in the 3+ state, whereas it is believed that Ce is in the 4+ state based on x-ray absorption spectroscopy measurements on a similar compound [11].

By comparison with the HTSC, the smaller Cu-O(1) bond length in $RuSr_2EuCeCu_2O_{10+\delta}$ would appear to imply that the hole concentration is greater than that in $RuSr_2Gd_{1.4}Ce_{0.6}Cu_2O_{10+\delta}$ or $RuSr_2GdCu_2O_8$. This is inconsistent with the absence of superconductivity in $RuSr_2EuCeCu_2O_{10+\delta}$. It is possible that the smaller Cu-O(1) bond length in $RuSr_2EuCeCu_2O_{10+\delta}$ when compared with $RuSr_2Gd_{1.4}Ce_{0.6}Cu_2O_{10+\delta}$ is due to the additional effect of the different $Ce^{4+}$ (ionic radii 0.97 Å) and $Gd^{3+}$ (ionic radii 1.053 Å) ionic radii. We note that a similar, but opposite, effect is observed in $TlSr_2(Lu_{1-x}Ca_x)Cu_2O_{7-\delta}$ where the substitution of $Ca^{2+}$ (ionic radii 1.12 Å) for $Lu^{3+}$ (ionic radii 0.977 Å) leads to an increase in the Cu-O(1) bond length [39].

There are small changes in the Cu-O-Cu buckling angle for $RuSr_2EuCeCu_2O_{10+\delta}$, $RuSr_2Gd_{1.4}Ce_{0.6}Cu_2O_{10+\delta}$ and $RuSr_2GdCu_2O_8$. However, while the average values are comparable to those found in $YBa_2Cu_3O_{7-\delta}$ (~166° [37]), they are clearly smaller than those found in the single $CuO_2$ layer superconductor, $La_{2-x}Sr_xCuO_4$ (180° [37]) at room temperature.

It has also been found that the Cu to in-plane oxygen bond length in the HTSC correlates with the hole concentration in the $CuO_2$ planes, where the Cu-O(2) bond length increases with decreasing hole concentration. Thus, the comparable Cu-O(2) bond lengths found in $RuSr_2Gd_{1.4}Ce_{0.6}Cu_2O_{10+\delta}$ and $RuSr_2GdCu_2O_8$ would seem to imply that these compounds have similar hole concentrations in the $CuO_2$ planes. This is consistent with the similar $T_c$ values where, for the HTSC, $T_c$ is found to correlate with the hole concentration [40]. Furthermore, the Cu-O(2) bond length is larger is $RuSr_2EuCeCu_2O_{10+\delta}$, which is consistent with this compound being more underdoped.

The Cu NQR spectra from $RuSr_2EuCeCu_2O_{10+\delta}$ are plotted in figure 7 above the magnetic ordering transition at 301.7 K (dashed curve) and 162.5 K (solid curve) as well as below the magnetic ordering transition at 102 K (dotted curve). The two peaks seen in the 162.5 K spectra near 31.2 MHz and 29.2 MHz are due to $^{63}Cu$ and $^{65}Cu$ nuclei in the $CuO_2$ planes respectively. The different frequencies for the $^{63}Cu$ and $^{65}Cu$ resonances arise from their different nuclear quadrupole moments and the relative intensities are due to their relative isotopic abundances. While the Cu NQR spectra are broad, we find that the NQR linewidth estimated from the higher frequency side of the $^{63}Cu$ NQR peak is ~2.2 MHz, which is comparable to that observed in $La_{2-x}Sr_xCuO_4$ [41,42]. However, the $^{63}Cu$ NQR linewidth in $La_{2-x}Sr_xCuO_4$ is anomalously broad when compared to other HTSC and it is independent of hole concentration for p>0.05. In the case of $La_{2-x}Sr_xCuO_4$, the broad NQR linewidths have been attributed to an inhomogeneous charge distribution [41]. It may be that the broad Cu NQR lines observed in the current $RuSr_2EuCeCu_2O_{10+\delta}$ sample are also due to an inhomogeneous charge distribution within the $CuO_2$ planes.

The Cu NQR spectra does not significantly broaden for temperatures down to 80 K, which is more than 30 K below the magnetic ordering temperature. Below 80 K the Cu NQR intensity decreases very rapidly and we were not able to record any spectra in the temperature region from 70 K down to 30 K. The Cu NQR signal was recovered for temperatures less that 30 K and in the same spectral range where the Cu NQR signal above 80 K was observed. However, the Cu NQR spectral width below 30 K exceeded the bandwidth of our experimental set-up (10 MHz). The broadening of the Cu NQR line at low temperatures, where the Ru moments are static on the NMR timescale, indicates that there is some hyperfine coupling from the Ru moments to Cu. The hyperfine field at the Cu site from Ru does not need to be large to significantly broaden the Cu NQR spectra. For example, a hyperfine contribution from Ru to the Cu site of only ~0.2 T, as reported for $RuSr_2GdCu_2O_8$ [13], could easily account for a broad spectral distribution of Cu NQR frequencies in zero field, especially as there is also magnetic disorder present due to ferromagnetic domain formation. We observe an enhancement of the Rabi precession frequency by a factor of 20, which is characteristic for ferromagnetic materials [43] and has also been observed in $RuSr_2GdCu_2O_8$ [13].

The room temperature $^{63}Cu$ NQR frequency is lower than that observed in $La_{2-x}Sr_xCuO_4$ (>31.5 MHz [42] and nearly temperature independent), slightly less than that observed in optimally doped YBa$_2$Cu$_3$O$_{7-\delta}$ (31.2 MHz at 300 K [44]) and clearly greater than that observed in HgBa$_2$CuO$_{4+\delta}$ (<24 MHz [45]). For these HTSC it has been found that there is a linear dependence of the Cu NQR frequency on hole concentration of the form, $\boldsymbol{\nu}_Q = a_1 p + a_0$, where $a_1$ and $a_0$ are positive and different for each family of HTSC. The variation of the Cu NQR frequency in the HTSC with increasing hole concentration is understood in terms of a changing electron density about the Cu nucleus [46].

We find that the $^{63}$Cu NQR frequency in RuSr$_2$EuCeCu$_2$O$_{10+\delta}$ increases by ~1 MHz when the temperature is decreased from room temperature to 80 K as can be seen in the insert to figure 7. A temperature dependent increase in the Cu NQR linewidth is observed in fully loaded YBa$_2$Cu$_3$O$_{7-\delta}$ and it has been attributed to lattice effects [44]. However, the temperature dependent increase observed in RuSr$_2$EuCeCu$_2$O$_{10+\delta}$ is anomously large and increases dramatically for temperatures less than ~200 K as can be seen in the insert to figure 7. The origin of this increase is not clear. We speculate that it may be due to structurally induced changes that result either in charge transfer to the CuO$_2$ layer or changes in the charge distribution at the planar Cu site [46].

We show in figure 8 that the spin-lattice relaxation magnetization recovery can be fitted to equation (1) at, and above, room temperature (open circles). Here we plot $F(\boldsymbol{t}) = (M(\boldsymbol{t}) - M_0)/2M_0$ (dashed curve) at 301.7 K and at 31.2 MHz. However, we find that the spin-lattice magnetization recovery at lower temperatures can not be fitted to a single exponential, as is apparent in figure 8 where we plot the spin-lattice magnetization recovery at 162.5 K and 31.2 MHz (crosses). The appearance of a spin-lattice magnetization recovery that can not be fitted to a single exponential indicates that there is a distribution of spin-lattice relaxation rates.

The distribution of $1/^{63}T_1$ can not be due to impurity phases because the synchrotron x-ray diffraction data indicates that the sample is single phase. In the absence of a specific spin-lattice relaxation rate distribution function, we fit the spin-lattice magnetization recovery to the stretched exponential function used by other researchers [47-49]. This function can be written as $F(\boldsymbol{t}) = \exp(-(3\boldsymbol{t}/^{63}T_1)^n)$, where $n$ is a measure of the $1/^{63}T_1$ distribution. We note that the $^{63}T_1$ obtained by the stretched exponential fit function gives an average spin-lattice relaxation time. We show it figure 8 that this stretched exponential function does fit the spin-lattice magnetization recovery at 162.5 K (solid curve).

We plot in figure 9 the resultant $n$ and $^{63}T_1 T$ values at 31.2 MHz (filled circles) where it can be seen that $n$ and $^{63}T_1 T$ decrease with decreasing temperature. A decrease in $^{63}T_1 T$ with decreasing temperature is also observed in optimally and overdoped YBa$_2$Cu$_3$O$_{7-\delta}$ and $La_{2-x}Sr_xCuO_4$, over the same temperature range. The expected behavior of $^{63}T_1T$ for metallic and magnetic systems, which was originally derived by Moriya, is given by [51]

$$(T_1T)^{-1} = \frac{1}{2}\hbar k_B \, g_n^2 \sum_{\mathbf{q}} |A(\mathbf{q})|^2 \, \frac{c''(\mathbf{q}, w_0)}{\hbar w_0}, \qquad (2)$$

where $|A(\mathbf{q})|$ is the form factor containing onsite and transferred hyperfine coupling constants, $\gamma_n$ is the nuclear gyromagnetic ratio, and $c''(\mathbf{q}, w_0)$ is the imaginary part of the dynamical spin susceptibility at the nuclear quadrupole resonance frequency, $\omega_0$. For HTSC, it is assumed that the main contribution to $|A(\mathbf{q})|$ for Cu arises from onsite hyperfine coupling as well as transferred hyperfine coupling to the four nearest-neighbor Cu sites. There is evidence of antiferromagnetic correlations in the HTSC and hence $c''(\mathbf{q}, w_0)$ has been modeled using the phenomenological dynamical spin susceptibility of Millis, Monien and Pines [52]. Consequently, the decrease in $^{63}T_1T$ with decreasing temperature observed in HTSC is attributed to changes in the antiferromagnetic spin fluctuation spectrum in the $CuO_2$ layers. The appearance of the normal-state pseudogap in underdoped HTSC causes $^{63}T_1T$ to begin to increase below a characteristic temperature.

Any significant additional hyperfine and dipolar coupling from the magnetic fluctuations in the $RuO_2$ layers to Cu in the $CuO_2$ layers should be evident by a rapid decrease of $^{63}T_1T$ as the magnetic ordering temperature (115 K) is approached. The functional form of the temperature dependence for this decrease in itinerant magnetic systems is given by $T_1T \propto (T^2 - T_M^2)^{1/2}$ or $T_1T \propto (T - T_M)^{1/2}$ for the random phase approximation or the renormalization spin fluctuation approaches, respectively, where $T_M$ is the magnetic ordering temperature [53] (115 K in $RuSr_2EuCeCu_2O_{10+\delta}$). We note that a significant decrease in $^{17}T_1T$ with decreasing temperature has been observed in $RuSr_2EuCu_2O_8$ from $^{17}O$ NMR measurements [54], where $^{17}T_1$ is the $^{17}O$ spin-lattice relaxation time. However, it is apparent in figure 9b that there is no dramatic divergence in $^{63}T_1T$ for $RuSr_2EuCeCu_2O_{10+\delta}$ near the magnetic ordering temperature, as would be expected by the preceding formulae. Rather, the data can be fitted to $^{63}T_1T \propto (T - T_0)$ for temperatures $\geq 80$ K. We note that correcting $^{63}T_1T$ for the increase in the Cu NQR frequency with decreasing temperature does not alter the linear relationship between $^{63}T_1T$ and temperature as can be seen in figure 9b (plus symbols).

A linear decrease in $^{63}T_1T$ with decreasing temperature is also observed in $La_{2-x}Sr_xCuO_4$ over the same temperature range [42]. Furthermore, the absolute values of $^{63}T_1T$ are within the range of those observed in $La_{2-x}Sr_xCuO_4$ is can be seen by the lines in figure 10b for 0.075 Sr (dotted curve [42]) and 0.24 Sr (dashed curve [42]). However, they are significantly less that those observed in $YBa_2Cu_3O_{7-\delta}$ (~0.25 sK at 300 K [35,42]). The $^{63}T_1T$ data in $RuSr_2EuCeCu_2O_{10+\delta}$ is complicated by the fact that the spin-lattice magnetization recovery can not be fitted to a single exponential recovery below room temperature as mentioned above and evident in figures 8 and 9a. Furthermore, it is apparent in figure 7 that $1/^{63}T_1$ also varies across the $^{63}Cu$ NQR line. One study on $La_{2-x}Sr_xCuO_4$ also found that the spin-lattice magnetization could not be fitted to a single exponential below 300 K [42]. However, this was not reported by two other studies [41,50]. A systematic decrease of $1/^{63}T_1$ with increasing frequency has been reported in $La_{2-x}Sr_xCuO_4$ in the same temperature region and it has been suggested that this is due to charge inhomogeneity [41]. This variation disappears on the overdoped side (p>0.16). It is unlikely that the behavior mentioned above is due to a slowing down of spin or charge fluctuations in the $CuO_2$ layers reported by other studies on $La_{2-x}Sr_xCuO_4$ because these effects are only expected to be observed below 100 K [47,51,55].

The simplest explanation for the Cu NQR data plotted in figures 7, 8 and 9 is that $^{63}T_1T$ in $RuSr_2EuCeCu_2O_{10+\delta}$ is dominated by hyperfine coupling within the $CuO_2$ layers and any additional hyperfine contribution from the spin fluctuations in the $RuO_2$ layers is small and therefore masked by the former. This interpretation is consistent with the known hyperfine coupling constants within the $CuO_2$ planes (37 T [56]) and if we assume that, like $RuSr_2GdCu_2O_8$ [13], the hyperfine coupling from Ru to Cu has an upper limit of 0.2 T. Thus, using equation (2) it can be shown that $1/^{63}T_1T$ is dominated by $CuO_2$ in-plane hyperfine coupling and the contribution from Ru to Cu hyperfine coupling contributes less than 0.01% to the total $1/^{63}T_1T$. Any additional relaxation mediated by dipolar coupling between the Ru moments and the Cu nuclei is also small, since the dipolar field from the Ru moments at the Cu site is estimated to be less than 1 T, where we use the structural data of tables 1 and 2 for the Ru-Cu distance and a Ru moment of 3 $\mu_B$.

The spin dynamics in the $CuO_2$ layers in $RuSr_2EuCeCu_2O_{10+\delta}$ is similar to those in the single $CuO_2$ layer HTSC $La_{2-x}Sr_xCuO_4$ rather than the double layer $YBa_2Cu_3O_{7-\delta}$. The variation in $1/^{63}T_1$ across the NQR line may indicate a local charge inhomogeneity, however the origin of the inhomogeneity may be different from that observed in $La_{2-x}Sr_xCuO_4$. We note that $La_{2-x}Sr_xCuO_4$ also has additional $Cu_A$ and $Cu_B$ sites which are believed to be due to Cu sites with Cu-O-La or Cu-O-Sr apical oxygen bonds [42]. This is not expected in $RuSr_2EuCeCu_2O_{10+\delta}$ because the apical oxygen bond is Cu-O-Ru and Ru has a fixed valence near 5.

**Conclusion**

In conclusion, we find that $RuSr_2EuCeCu_2O_{10+\delta}$ ferromagnetically orders with a single transition temperature of 115 K. The magnetic ordering temperature is lower than that observed in the antiferromagnetic superconductors, $RuSr_2GdCu_2O_8$ and $RuSr_2EuCu_2O_8$. It is clear that $RuSr_2EuCeCu_2O_{10+\delta}$ has a sizeable ferromagnetic component and the moment per Ru at 6 T is remarkable similar to that observed in $RuSr_2EuCu_2O_8$ at 6 T. Furthermore, $RuSr_2EuCeCu_2O_{10+\delta}$ behaves like a soft ferromagnet with most of the rapid ordering of the Ru moments being completed after ~0.2 T. However, the deduced Ru moment above the magnetic ordering temperature is significantly less that that expected for a S=3/2, $t_{2g}^3$ ground state. The comparable tilt and rotation angles found in $RuSr_2EuCeCu_2O_{10+\delta}$, $RuSr_2Gd_{1.4}Ce_{0.6}Cu_2O_{10+\delta}$ and $RuSr_2GdCu_2O_8$ indicate that the structural mechanism responsible for changes in the magnetic order found in $Sr_{1-x}Ca_xRuO_3$ can not be applied to the ruthenate-cuprates. We find a distribution of $1/^{63}T_1$ below 300 K and $1/^{63}T_1$ varies across the $^{63}Cu$ NQR line. Furthermore, the Cu NQR data are remarkable similar to $La_{2-x}Sr_xCuO_4$ in the metallic state and $1/^{63}T_1$ is dominated by hyperfine coupling in the $CuO_2$ layers where any additional hyperfine coupling from the spin fluctuations in the $RuO_2$ layers is small. The functional form and frequency dependence of $1/^{63}T_1$ gives some evidence for an intrinsic inhomogeneous local charge distribution within the $CuO_2$ planes of this material.

**Acknowledgements**

We acknowledge funding support from the New Zealand Marsden Fund (GVMW), the Alexander von Humboldt Foundation (GVMW), and from the Kangwon National University, support of 2000 faculty research abroad (H. K. Lee). We thank EPSRC for the provision of research grant GR/M59976, Daresbury SRS beam time and a studentship for ACM, M. A. Roberts and H. Nowell for help with the synchrotron diffraction experiment. We acknowledge helpful discussions with J. Haase.

**TABLES**

**Table 1:** Refined atomic parameters for $RuSr_2(Eu_{2-x}Ce_x)Cu_2O_{10+\delta}$ solid solutions from room temperature synchrotron X-ray diffraction data. The space group is I4/mmm and the atom positions are Ru 2(a) (0, 0, 0), Sr 4(e) (½, ½, z), Eu/Ce 4(e) (½,½, z), Cu 4(e) (0, 0, z), O(1) 16(n) (x, 0, z), O(2) 8(g) (0, ½, z), O(3) 8(j) (x, ½, 0), O(4) 4(d) (0, ½, ¼).

| *Atom* | Occupancy | | Ru1222:1.0Eu |
|---|---|---|---|
| Ru | 1.00 | $U_{iso}$ (Å$^2$) | 0.0033(4) |
| Sr | 1.00 | z | 0.07743(5) |
| | | $U_{iso}$ (Å$^2$) | 0.0120(4) |
| Eu/Ce | 1.00 | z | 0.20478(2) |
| | | $U_{iso}$ (Å$^2$) | 0.0002(4) |
| Cu | 1.00 | z | 0.14351(6) |
| | | $U_{iso}$ (Å$^2$) | 0.00019(2) |
| O(1) | 0.25 | x | 0.052(5) |
| | | z | 0.0691(3) |
| | | $U_{iso}$ (Å$^2$) | 0.002(4) |
| O(2) | 1.00 | z | 0.1497(3) |
| | | $U_{iso}$ (Å$^2$) | 0.008(1) |
| O(3) | 0.50 | x | 0.127(3) |
| | | $U_{iso}$ (Å$^2$) | 0.005(2) |
| O(4) | 1.00 | $U_{iso}$ (Å$^2$) | 0.002(1) |

**Table 2:** Refined cell parameters, agreement factors and selected bond lengths and angles for the RuSr$_2$EuCeCu$_2$O$_{10+\delta}$ solid solutions. Also shown is data for RuSr$_2$Gd$_{1.4}$Ce$_{0.6}$Cu$_2$O$_{10+\delta}$ [19] and RuSr$_2$GdCu$_2$O$_8$ [18].

| | Ru1222:1.0Eu | Ru1222:1.4Gd | Ru1212:Gd |
|---|---|---|---|
| **Eu/Ce-O(2)** | 2.485(5) | | |
| **Eu/Ce-O(4)** | 2.3166 (4) | | |
| **Cu-O(1) (Å) X 1** | 2.135(9) | 2.213(4) | 2.184(6) |
| **Cu-O(2) (Å) X 4** | 1.9309(7) | 1.9251(3) | 1.9268(4) |
| **Sr-O(1) (Å) X 2** | 2.59(1) | | |
| **Sr-O(1) (Å) X 2** | 2.88(1) | | |
| **Sr-O(2) (Å) X 4** | 2.822(6) | | |
| **Sr-O(3) (Å) X 2** | 2.637(6) | | |
| **Sr-O(3) (Å) X 2** | 3.273(8) | | |
| **Ru-O(1) (Å) X 2** | 1.986(9) | 1.9298(3) | 1.933(5) |
| **Ru-O(3) (Å) X 4** | 1.984(3) | 1.975(1) | 1.969(2) |
| **O(1)-Cu-O(1) (°)** | 89.8-100.7 (5) | | |
| **Cu-O(1)-Ru (°)** | 168.7(9) | 168.2(4) | 171.5(8) |
| **Cu-O(2)-Cu (°)** | 169.5(4) | 169.2(2) | 169.9(2) |
| **Ru-O(3)-Ru (°)** | 151.5(6) | 153.2(2) | 154.3(4) |
| **a (Å)** | 3.84554 (2) | 3.8423(10) | 3.83032(3) |
| **c (Å)** | 28.5760(2) | 28.5803(9) | |
| **v (Å$^3$)** | 422.587(7) | | |
| **c$^2$** | 2.93 | | |
| **R$_{WP}$** | 7.77 | | |
| **R$_P$** | 6.06 | | |

**FIGURES**

**Figure 1:** Unit cell of $RuSr_2EuCeCu_2O_{10+\delta}$. The tilting and rotation of the $RuO_6$ octahedra are also shown.

**Figure 2:** (a) Real part of the ac susceptibility for $RuSr_2EuCeCu_2O_{10+\delta}$ (solid curve) and $RuSr_2EuCu_2O_8$ (dashed curve). (b) Imaginary part of the ac susceptibility for $RuSr_2EuCeCu_2O_{10+\delta}$ (solid curve).

**Figure 3:** Plot of the magnetization against temperature for $RuSr_2EuCeCu_2O_{10+\delta}$ (solid curve and left axis) and $RuSr_2EuCu_2O_8$ (dashed curve and right axis). The lower curves are zero-field-cooled and the upper curves are field-cooled with an applied magnetic field of –2.5 mT. Upper insert: Plot of the difference between the field-cooled and zero-field-cooled magnetization curves for $RuSr_2EuCeCu_2O_{10+\delta}$ (solid curve and left axis) and $RuSr_2EuCu_2O_8$ (dashed curve and right axis). Lower insert: Plot of the magnetic ordering temperature in $RuSr_2EuCeCu_2O_{10+\delta}$ as defined as the temperature where the maximum negative value of $dM/dT$ occurs.

**Figure 4:** Plot of the magnetization against applied magnetic field for $RuSr_2EuCeCu_2O_{10+\delta}$ (solid curve) and $RuSr_2EuCu_2O_8$ (dashed curve). Insert: Plot of the magnetization over the full measured magnetic field range for $RuSr_2EuCeCu_2O_{10+\delta}$ (solid upper curve) and $RuSr_2EuCu_2O_8$ (dashed lower curve).

**Figure 5:** Plot of the dc magnetization for an applied magnetic field of 1 T, 3 T and 6 T (upper solid curves) $RuSr_2EuCeCu_2O_{10+\delta}$. The arrow indicates increasing magnetic field. Also included is the estimated $Eu^{3+}$ contribution (dotted curve), the dc magnetization at 1 T after subtraction of the $Eu^{3+}$ contribution (dashed curve) and the best fit to the equation described in the text (solid curve).

**Figure 6:** Plot of the synchrotron x-ray spectra from $RuSr_2EuCeCu_2O_{10+\delta}$. Also shown is the difference spectra after fitting the spectra to the I4/mmm space group and with the refined atomic parameters in table 1.

**Figure 7:** Plot of the Cu NQR spectra at 301.7 K (dashed curve). 162.5 K (solid curve) and 102 K (dotted curve). Also shown is $1/^{63}T_1$ at 301.7 K (open circles), 162.5 K (crosses) and 102 K (filled circles). Insert: Plot of the $^{63}Cu$ NQR peak frequency against temperature.

**Figure 8:** Plot of the Cu NQR spin-lattice decay function against temperature at 162.5 K (crosses) and 301.7 K (open circles). The curves are best fits to the data using a stretched exponential (solid curve) and a monoexponetial (dashed curve).

**Figure 9:** (a) Plot of the stretched exponential exponent, $n$, against temperature at 31.2 MHz. (b) plot of $^{63}T_1T$ at 31.2 MHz (filled circles). Also shown is $^{63}T_1T$ at the $^{63}$Cu NQR peak frequency (plus symbols) and $^{63}T_1T$ from $La_{2-x}Sr_xCuO_4$ with 0.24 Sr (dashed curve [42]) and 0.075 Sr (dotted curve [42]). The solid line is a linear fit to the data. The vertical arrow indicates the magnetic ordering temperature of 115 K.

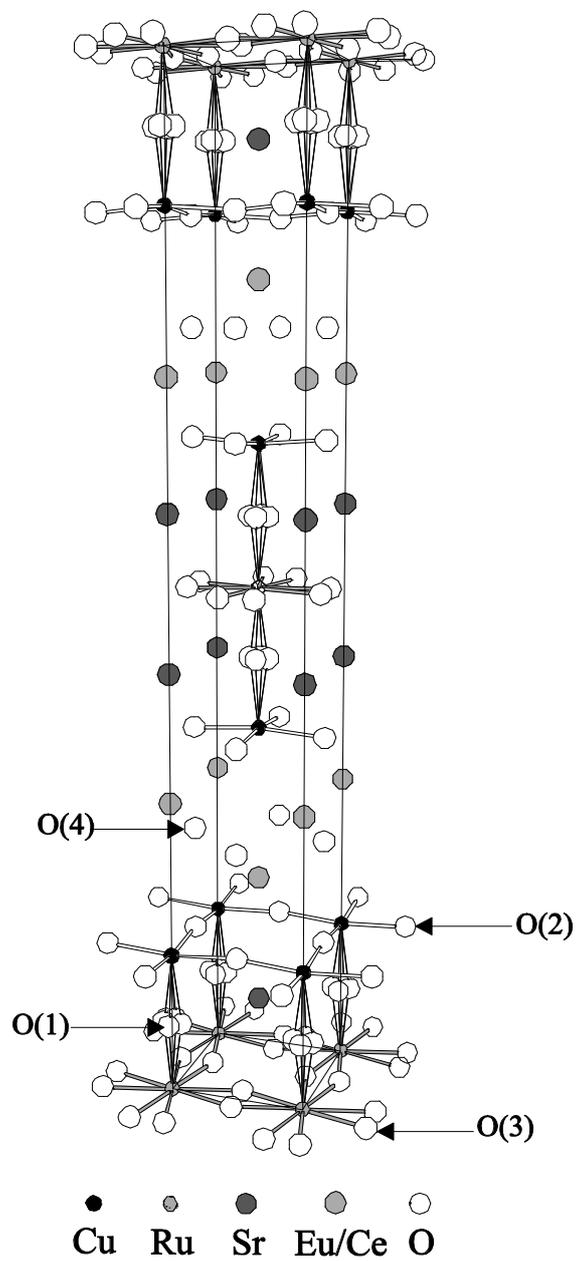

Figure 1
Williams *et al.*
Phys. Rev. B

Cu   Ru   Sr   Eu/Ce   O

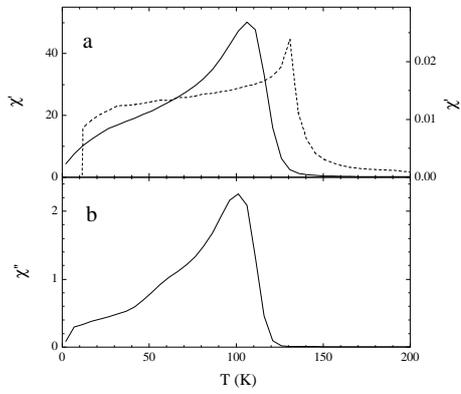

Figure 2
Williams *et al.*
Phys. Rev. B

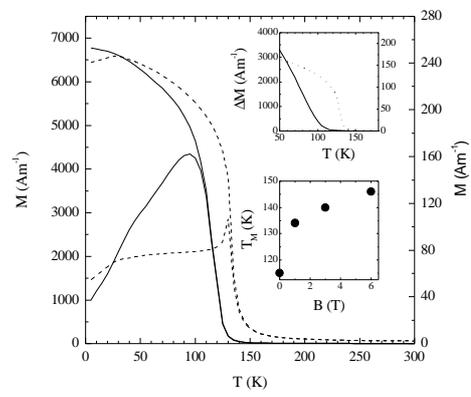

Figure 3
Williams *et al.*
Phys. Rev. B

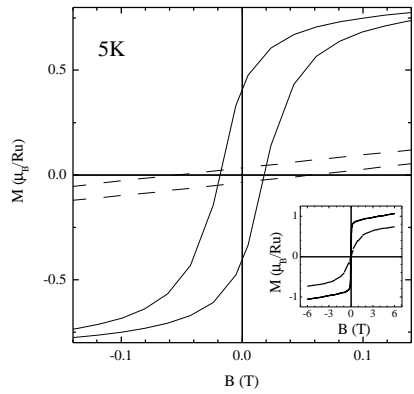

Figure 4
Williams *et al.*
Phys. Rev. B

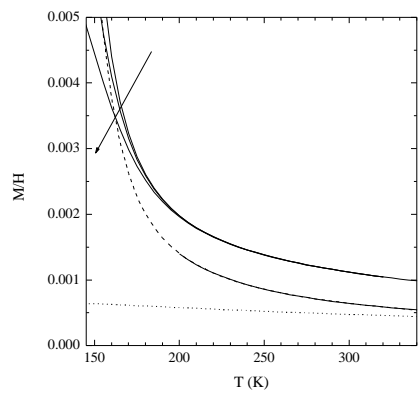

Figure 5
Williams *et al.*
Phys. Rev. B

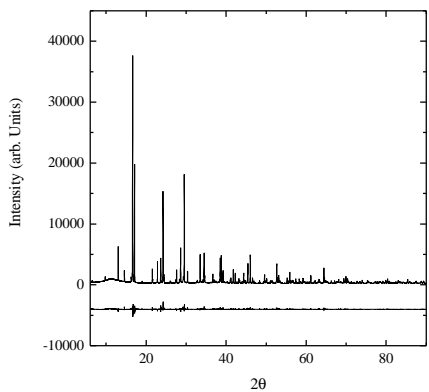

Figure 6
Williams *et al.*
Phys. Rev. B

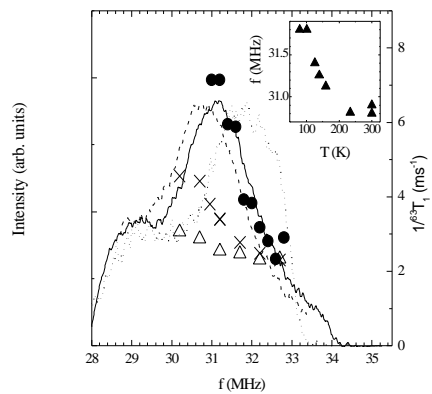

Figure 7
Williams *et al.*
Phys. Rev. B

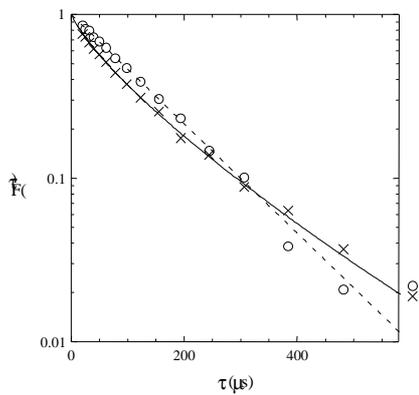

Figure 8
Williams *et al.*
Phys. Rev. B

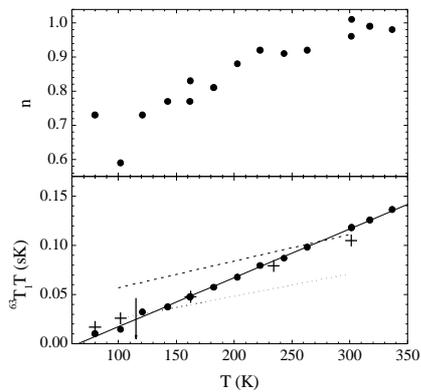

Figure 9
Williams *et al.*
Phys. Rev. B